\documentclass[fleqn,usenatbib]{mnras}

\usepackage{newtxtext,newtxmath}

\usepackage[T1]{fontenc}
\usepackage{ae,aecompl}

\usepackage{graphicx}	
\usepackage{amsmath}	
\usepackage{amssymb}	

\title[Binary Black Holes in Open Clusters]{Gravitational-Wave Emission from Binary Black Holes Formed in Open Clusters}

\author[J. Kumamoto et al.]{
Jun Kumamoto,$^{1}$ \thanks{E-mail:kumamoto@astron.s.u-tokyo.ac.jp}
Michiko S. Fujii,$^{1}$ 
and Ataru Tanikawa,$^{2,3}$
\\
$^{1}$ Department of Astronomy, Graduate School of Science, The University of Tokyo, 7-3-1 Hongo, Bunkyo-ku, Tokyo 113-0033, Japan\\
$^{2}$ Department of Earth Science and Astronomy, College of Arts and Sciences, The University of Tokyo, 3-8-1 Komaba, Meguro-ku, Tokyo 153-8902, Japan \\
$^{3}$ RIKEN Center for Computational Science, 7-1-26 Minatojima-minami-machi, Chuo-ku, Kobe, Hyogo 650-0047, Japan
}

\date{Accepted XXX. Received YYY; in original form ZZZ}

\pubyear{XXXX}

\begin{document}
\label{firstpage}
\pagerange{\pageref{firstpage}--\pageref{lastpage}}
\maketitle

\begin{abstract}
In order to investigate the formation rate of binary black holes (BBHs) in stellar clusters with a mass comparable to open clusters, we performed a series of direct $N$-body simulations of open clusters with a mass of $2.5\times10^3$ (Model A) and $10^4 M_{\odot}$ (Model B). Since such low-mass clusters would have been more populous than globular clusters when they were born, low-mass clusters are also candidates as the origin of BBHs which are the source of the gravitational waves. In model A, most of BBHs merged within 10\,Gyr formed via dynamically formed main-sequence binary stars and stable and unstable mass transfer between them since open clusters collapse within the main-sequence life-time of massive stars. These binaries, therefore, have little eccentricities. The fraction of such binaries among all merging BBHs increases as the cluster mass decrease due to the shorter relaxation time. In our simulations, $4.0\times 10^{-5}$ and $1.7\times10^{-5}$ BBHs per solar mass merged within 10\,Gyr for models A and B, respectively. These values correspond to $\sim 20$--50\,\% of the number of mergers per solar mass originated from globular clusters with a mass of $10^5$--$10^6M_{\odot}$. Thus, the contribution of BBHs originated from open clusters is not negligible. The estimated mergers rate density in the local universe is about 0.3 \,yr$^{-1}$\,Gpc$^{-3}$ assuming a cluster mass function with a power of $-2$.
\end{abstract}





\section{Introduction}

The first gravitational wave direct detection was presented by LIGO \citep{2016PhRvL.116f1102A}. After the first detection, BBH mergers has been observed during the observation period \citep{2016PhRvL.116x1103A,2017PhRvL.118v1101A,2017PhRvL.119n1101A,2017ApJ...851L..35A}. These detections suggest that $>10$ solar mass black-holes binaries commonly exists in the universe. 

Two scenarios have been proposed for the origin of such a tight BBH. One is stable and unstable mass-transfer between tight stellar binaries \citep[e.g.][]{2014MNRAS.442.2963K,2016Natur.534..512B}, and the other is the dynamical formation due to the three-body encounters in the dense core of star clusters \citep{2000ApJ...528L..17P}. The sites of the dynamical formation could also be galactic nuclei \citep{2009MNRAS.395.2127O,2016ApJ...831..187A}.

For the dense cluster scenario, globular clusters with a mass of $10^5$--$10^6M_{\odot}$ has mainly been focused on in previous studies \citep{2000ApJ...528L..17P,2006ApJ...637..937O,2008ApJ...676.1162S,2010MNRAS.407.1946D,2011MNRAS.416..133D,2010MNRAS.402..371B,2013MNRAS.435.1358T,2014MNRAS.440.2714B,2015PhRvL.115e1101R,2016PhRvD..93h4029R,2017PASJ...69...94F,2017MNRAS.469.4665P,2017MNRAS.464L..36A,2018MNRAS.480.5645H}. Since globular clusters typically has very dense cores ($>10^4M_{\odot}\,{\rm pc}^{-3}$) \citep{2008MNRAS.389.1858H,2009MNRAS.395.1173G} and therefore can form merging BBHs comparable to the observed merger rate density.

On the other hand, less massive clusters with a mass of $10^{3-4}M_{\odot}$ has not been expected as the origin of BBHs because only a few massive BHs can form in one cluster. In addition, the gravitational potential of these clusters are much shallower than typical globular clusters, and therefore it seems to be difficult to form tight binaries. Actually, only a few previous studies have investigated the formation of BBHs in star clusters with a mass of $10^4M_{\odot}$ \citep{2017MNRAS.467..524B,2018MNRAS.473..909B,2018MNRAS.481.5123B}. In these studies, however, a few merging BBHs formed in one cluster.

The number of star clusters with a mass comparable to open clusters would have been an order of magnitude larger than that of globular clusters when they formed.
Observations of young star clusters in the Milky Way and nearby galaxies suggests that the initial mass distribution of star clusters has a power of $-2$ \citep[][and references therein]{2010ARA&A..48..431P}. Not only the observation of nearby young star clusters, but also the observed mass distributions of globular clusters and tidal disruption scenario also suggested that the number of smaller clusters increases following a power of $-2$ or similar \citep{FallZhang2001}.
If we assume that the number of black holes simply correlated 
with the total stellar mass and the cluster mass function follows $m^{-2}$, the total number of black holes per solar mass for $10^{3-4}M_{\odot}$ clusters should be the same as that for $10^{5-6}M_{\odot}$ clusters. 

In this paper, we investigate the formation rate of BBHs in star clusters with a mass range of $10^3$ to $10^4M_{\odot}$, which is less massive than previous studies, using direct $N$-body simulations. We show a new channel for the formation of BBHs which is dominant in open clusters.

\section{Methods and Models}

\subsection{Initial conditions}
We simulated two cluster models (A and B) of which the initial masses ($M_{\rm cl,ini}$) are $2.5\times10^3$ and $1.0\times10^4M_{\odot}$, respectively.
We adopt Plummer profile \citep{1911MNRAS..71..460P} as the initial density profile;
\begin{equation}
    \rho(r) = \frac{3M}{4\pi a^3} \left( 1+\frac{r^2}{r_p^2} \right)^{-5/2},
	\label{eq:plummer}
\end{equation}
where $r_p$ is a radial scale parameter.
In Table \ref{tab:models}, we summarize the parameters of our models.
In this profile, the half mass radius $r_{\rm hm}$ is given by
\begin{equation}
    r_{\rm hm} = (2^{2/3}-1)^{-1/2}r_p.
	\label{eq:r_hm}
\end{equation}

We adopt the half-mass density, which is the mass density inside the half mass radius ($\rho_{\rm hm} = 3M/8\pi r_{\rm hm}^3$), as a parameter. We set the initial half-mass density to be $10^4M_{\odot}$\,pc$^{-3}$. This value is much larger than those of observed open clusters \citep{2010ARA&A..48..431P}, but an order of magnitude smaller than that of typical initial conditions for globular clusters \citep{2008MNRAS.389.1858H}. 
Even if we chose this initial density, the open clusters collapse before first supernova explosion (3--4\,Myr) and evolve to the current density of observed open clusters \citep{2016ApJ...817....4F}.

Once the mass and half-mass density are given, we can scale the distribution and calculate the half-mass relaxation time:
\begin{eqnarray}
    t_{\rm rh} &=& 0.0477 \frac{N}{(G\rho_{\rm hm})^{1/2}\log(0.4N)} \\
               &\sim& 0.711\frac{N}{\log(0.4N)}\left(\frac{\rho_{\rm hm}}{M_{\odot}{\rm pc^-3}}\right)^{-0.5}~{\rm Myr}.
	\label{eq:t_rh}
\end{eqnarray}
The half-mass relaxation time of our initial conditions ($t_{\rm rh, ini}$) are 9.38 and 31.6\,Myr for Models A and B, respectively. Star clusters with an equal-mass stars collapse after 15--20$t_{\rm rh, ini}$ \citep{1987degc.book.....S}, but the core-collapse time becomes shorter correlated with the mass ratio between the mean and most massive star \citep{2004ApJ...604..632G,2014MNRAS.439.1003F}. The expected core-collapse time is within 4\,Myr. These models, therefore, can reach core collapse before the first supernova explosion.

\begin{table*}
\caption{The initial parameters of our models.}
\centering
\begin{tabular}{lcccccc}
\hline
& $M_{\rm cl,ini} [M_\odot]$ & $N_{\rm ini}$ & $N_{\rm run}$ & $\rho_{\rm hm} [M_\odot {\rm pc^{-3}}]$ & $r_{\rm hm} [{\rm pc}]$ & $t_{\rm rh,ini} [{\rm Myr}]$ \\
\hline \hline
Model A & $2.5\times10^3$ &  4266 & 360 & $10^4$ & 0.310 & 9.38 \\
Model B & $1.0\times10^4$ & 17064 &  90 & $10^4$ & 0.492 & 31.6 \\
\hline
\end{tabular}
\label{tab:models}
\end{table*}

The initial mass of each particle is given randomly from the Kroupa initial mass function \citep{Kroupa2001}. We set the lower and upper limit of the particle mass to be $m_{\rm min} = 0.08 M_{\odot}$ and $m_{\rm max} = 150M_{\odot}$, respectively. With this set-up, the expected average mass of the stellar particles is $\langle m \rangle = 0.586M_{\odot}$. The initial number of particles in each model was given by
\begin{equation}
    N_{\rm ini} = \frac{M_{\rm cl,ini}}{\langle m \rangle}.
	\label{eq:Nini}
\end{equation}
The resulting number of particles ($N_{\rm ini}$) is 4266 and 17064 for models A and B, respectively. We did not assume primordial binaries. The dependence of the primordial binary fraction to the number of ejected BBHs is, however, given by an equation in \citet{2018MNRAS.480.5645H}. It is possible to estimate the case with primordial binaries using this equation.

We performed 360 and 90 runs for Model A and B with different random seeds. The total cluster mass of all runs is $9.0\times10^5M_{\odot}$ for both models.

\subsection{$N$-body simulations}

We use {\tt NBODY6++GPU} \citep{Wang+2015}, which is an MPI-parallelized and GPU enabled version of {\tt NBODY6} \citep{1999PASP..111.1333A}. The stellar orbits are integrated by fourth-order Hermite scheme \citep{1992PASJ...44..141M} and binaries are treated using KS regularization \citep{KustaanheimoStiefel1965, 1993CeMDA..57..439M}. We use the GPU cluster SGI Rackable C1102-GP8 (Reedbush-L) in the Information Technology Center, The University of Tokyo.

We did not assume the tidal effect. Our models collapsed in $<2$\,Myr and formed BBHs mostly within 100\,Myr. In such a short timescale, the effect of tidal disruption is relatively small. 
We performed simulation to 3\,Gyr.

\subsection{Stellar evolution}

The latest {\tt NBODY6} contains a stellar evolution model \citep{Hurley+2000} with an updated mass loss model \citep{Belczynski+2010}. However, {\tt NBODY6++GPU} does not contain the updated model. We transported the updated mass loss model from the latest {\tt NBODY6} to {\tt NBODY6++GPU} and use this model. The evolution of stellar radius, mass, luminosity and mass loss of each star can be calculated from the initial stellar mass and metallicity. We set the stellar metallicity to be $0.1$ solar metallicity.

We switched off the velocity kick caused by asymmetric supernovae explosion for fair comparison with previous studies such as \citet{2018MNRAS.473..909B}, in which no natal kicks for the neutron stars and direct-collapse BHs and kicks smaller than the escape velocity are assumed. The number of ejected BBHs is expected to be simply proportional to the ejection rate due to the kick \citep{2013MNRAS.435.1358T}.

\section{Results}

\subsection{Properties of Ejected Binary Black Holes}

\begin{figure}
\begin{center}
\includegraphics[width=0.45\textwidth]{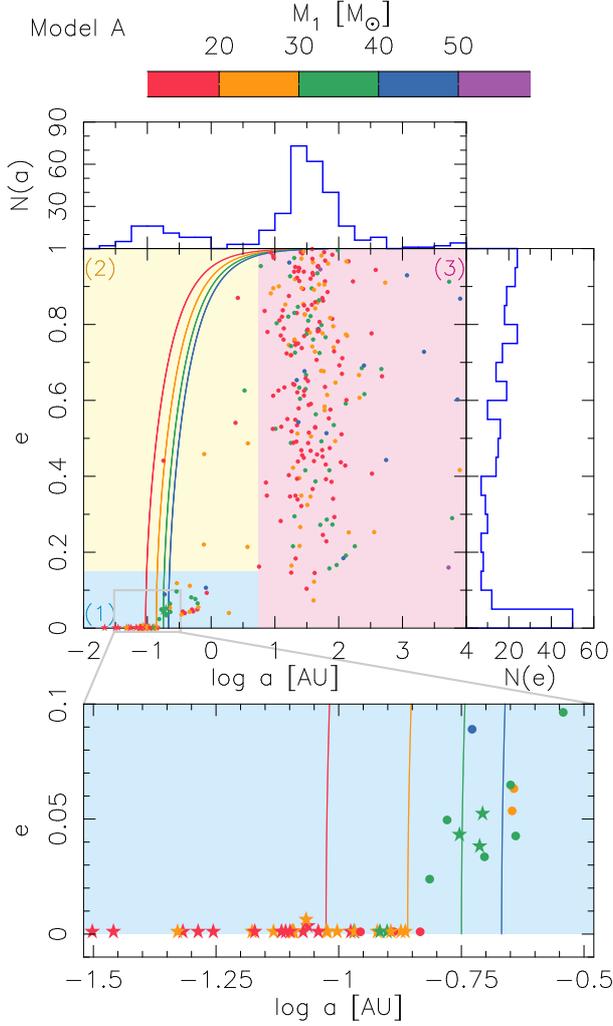}
\end{center}
\caption{The distribution of eccentricity and semi-major axis  of BBHs ejected from the star clusters for Models A. The color of each point shows a $M_1$ of BBH. The solid lines show $T_{\rm GW} = 10~{\rm Gyr}$ with $M_1 = 15$, $25$, $35$ and $45M_{\odot}$ in order from the left to right with $q=0.5$. The lower panel displays the close up view of a part of the upper panel.
}
\label{fig:a-e1}
\end{figure}

\begin{figure}
\begin{center}
\includegraphics[width=0.45\textwidth]{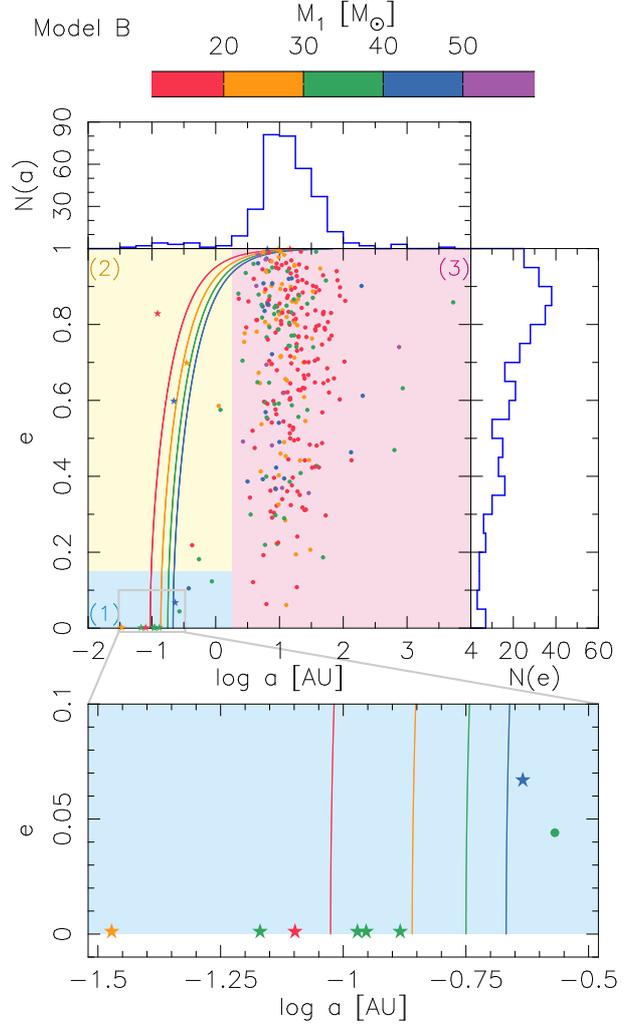}
\end{center}
\caption{Same as Figure \ref{fig:a-e1}, but for Model B.
}
\label{fig:a-e2}
\end{figure}

    In Figures \ref{fig:a-e1} and \ref{fig:a-e2}, we present the distribution of eccentricity ($e$) and semi-major axis ($a$) of BBHs ejected from our cluster models A and B, respectively. The color of each point shows the primary mass ($M_1$) of BBH. The solid lines show the boundary where the merger time is equal to $10~{\rm Gyr}$ when the $M_1$ is $15$, $25$, $35$ and $45M_{\odot}$ in order from the left to right with a mass ratio $q=0.5$. Therefore, BBHs existing to the left side from the solid line are expected to merged within the cosmic age. The BBHs with $T_{GW} < 10~{\rm Gyr}$ is showed with star symbols in Figures \ref{fig:a-e1} and \ref{fig:a-e2}. Detailed results on merger time are described in the next subsection.

Based on the distribution of BBHs in Figures \ref{fig:a-e1} and \ref{fig:a-e2}, we divided the $a$ -- $e$ space into three color shaded regions, Region 1, 2 and 3. For Model A, Region 1, 2 and 3 are defined by $\{ (a,e)~|~a \leq 10^{0.75}, e \leq 0.15 \}$, $\{ (a,e)~|~a \leq 10^{0.75}, 0.15 < e \}$ and $\{ (a,e)~|~10^{0.75} < a \}$, respectively. For Model B, Region 1, 2 and 3 are defined by $\{ (a,e)~|~a \leq 10^{1.25}, e \leq 0.15 \}$, $\{ (a,e)~|~a \leq 10^{1.25}, 0.15 < e \}$ and $\{ (a,e)~|~10^{1.25} < a \}$, respectively. 
The border between Region 2 and 3 is determined by the difference in binary formation. BBHs in Region 2 are evolved from binaries of massive main-sequence stars or giants via stable and unstable mass transfer evolution. On the other hand, BBHs in Region 3 are formed via three-body encounter of BHs. This border is not exact, because the formation path is not clearly separated only by the orbital elements when the BBHs are ejected.

\begin{figure*}
\begin{center}
\includegraphics[width=0.90\textwidth]{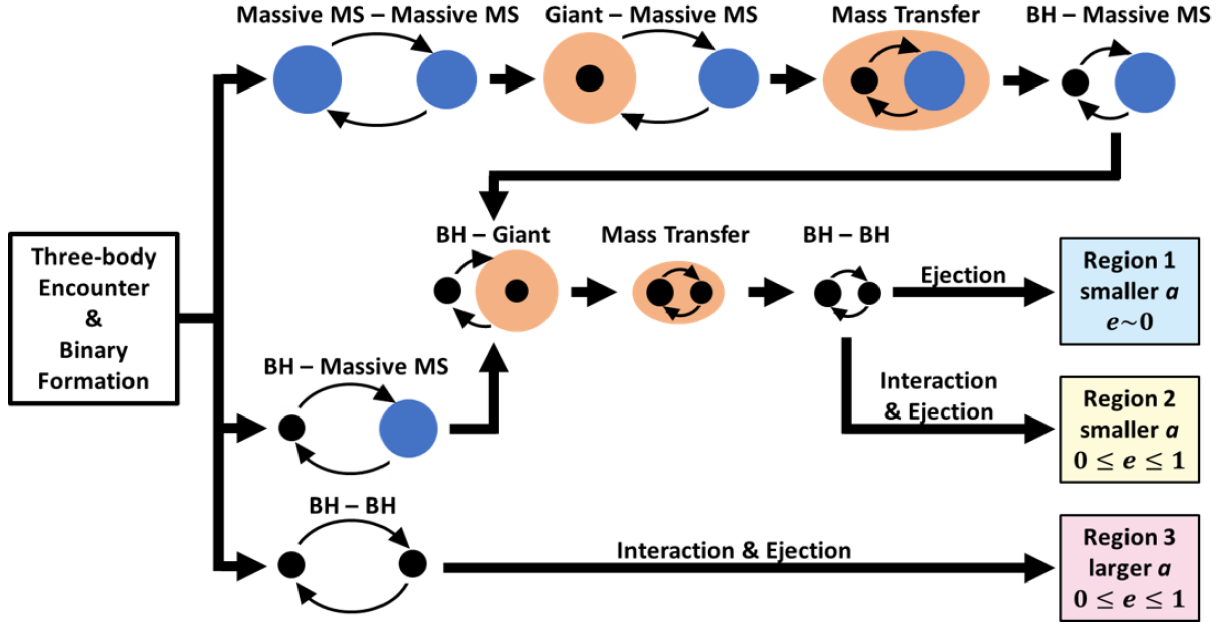}
\end{center}
\caption{The schematic illustration of the formation process of BBH in our simulation. Depending on the formation process, the final features of BBHs are classified into three groups shown by color shaded regions in Figures \ref{fig:a-e1} and \ref{fig:a-e2}. MS in figure stands for main-sequence.
}
\label{fig:BinEvo}
\end{figure*}

In the blue shaded region (Region 1) in the Figures \ref{fig:a-e1} and \ref{fig:a-e2}, there are some harder BBHs. Several of such BBHs have shorter merger time than $10~{\rm Gyr}$. These BBHs formed binaries during their main-sequence life-time, and therefore they experienced common envelope evolution. As a result, they have nearly circular orbits. Since these BBHs are very tight, they behave as single stars. They interact with other softer binaries and are ejected conserving their very small eccentricity. 

The fraction of such binaries increases as the cluster mass decrease. This is because the relaxation time decreases as the cluster mass decreases, if we fixed the half-mass density of the clusters. For the formation of these binaries, massive stars must form binaries before they evolve to BHs, i.e., the cluster have to reach core collapse before massive stars evolve to BHs.

For stars with a mass of $100M_{\odot}$, the main-sequence life-time is $\sim 3.5~{\rm Myr}$ \citep{Hurley+2000}. The core-collapse time of star clusters with a mass function is estimated by
\begin{equation}
  t_{\rm cc} \sim 0.07 t_{\rm rh,ini},
\end{equation}
for the cluster with $m_{\rm max} / \langle m \rangle > 50$ \citep{2004ApJ...604..632G,2016ApJ...817....4F}. From this estimation, the core-collapse time for our Models A and B are equal to $\sim 0.7~{\rm Myr}$ and $\sim 2~{\rm Myr}$, respectively. Therefore, the core collapse can occur before massive BHs formation for our cluster models.

If such BBHs are not tight enough, they behave as a binary. They experience three-body encounters and change their eccentricities. These stars are located in the yellow shaded region (Region 2) in Figures \ref{fig:a-e1} and \ref{fig:a-e2}. 

Most of the ejected binaries distribute the red shaded region (region 3) in the figures. These binaries formed via three-body encounters of BHs. Their eccentricity distribution is similar to the thermal distribution, although there are statical fluctuations due to the limited number of runs. The peak semi-major axis of binaries in this region shifts to smaller as the cluster mass increases. This can be explained as follows. In a cluster, a BBH becomes tighter through interactions with other cluster stars. As a backreaction of the tightening, the BBH itself and interacting stars get kick velocity. When the kick velocity is larger than the escape velocity of a cluster, the BBH is ejected from the cluster. Hence, a BBH tends to become tighter in a cluster with larger escape velocity or larger mass.

\subsection{Black hole mergers within a Hubble time}
\begin{figure}
\begin{center}
\includegraphics[width=0.45\textwidth]{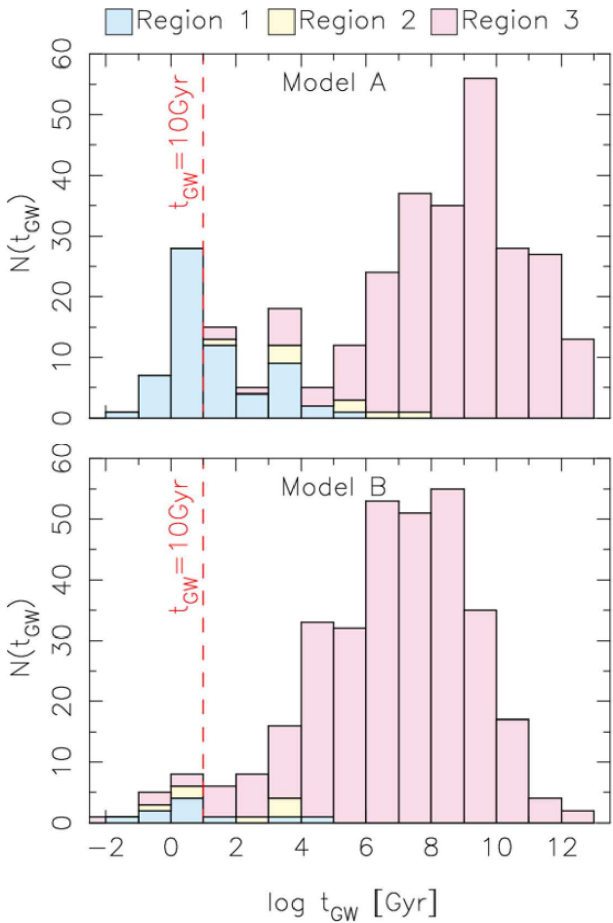}
\end{center}
\caption{The distribution of the merging time ejected BBHs in each Region for Models A (top pnael) and B (bottom panel). The red dashed line shows that $t_{\rm GW} = 10~{\rm Gyr}$.
}
\label{fig:tgw}
\end{figure}

We calculate merger time using the equation \citep{1963PhRv..131..435P}:
\begin{eqnarray}
    t_{\rm GW} &=& \frac{5}{256} \frac{c^5}{G^3} \frac{a^4}{M_1^3q(1+q)} g(e) \\
    &\sim& 1.2\times10^4 \left( \frac{M_1}{30M_{\odot}} \right)^{-3} \left( \frac{a}{1~{\rm AU}} \right)^{4} \frac{g(e)}{q(1+q)}~{\rm Gyr},
	\label{eq:tgw}
\end{eqnarray}
where
\begin{equation}
    g(e) = \frac{(1-e^2)^{3.5}}{1+(73/24)e^2+(37/96)e^4}.
\end{equation}
Here, $c$, $G$ and $q$ are light speed, gravitational constant and mass ratio of BBH, respectively. Figure \ref{fig:tgw} shows the distribution of merger time of ejected BBHs in our simulations.

\begin{table*}
\caption{The number of binary black holes}
\centering
\begin{tabular}{lccc}
\hline
& Model A & Model B & \citet{2016PhRvD..93h4029R}\\
\hline \hline
$M_{\rm cl,ini}[M_{\odot}]$ & $2.5\times 10^3$& $1.0\times10^4$ & $6\times10^5$\\
$N_{\rm BBH,esc}$ per $9\times 10^5 M_{\odot}$ & 323 & 335 & 182--210\\
$N_{\rm BBH,esc} (t_{\rm GW}<10~{\rm Gyr})$ per $9\times 10^5M_{\odot}$ & 36 & 15 & 65--81 \\
-- in Region 1 & 36 & 7 & \\
-- in Region 2 & 0 & 3 & \\
-- in Region 3 & 0 & 5 & \\
Mergers per solar mass [$M_{\odot}^{-1}$] & $4.0\times10^{-5}$ & $1.7\times10^{-5}$ & 7.2--$9.2\times10^{-5}$\\
\hline
\end{tabular}
\newline
{ \scriptsize
For \citet{2016PhRvD..93h4029R}, we choose models with a half-mass density comparable to our model ($10^4M_{\odot}$pc$^{-3}$), and then, estimate the number of ejected binary black holes from the minimum and maximum value of three different metallicity models (0.01, 0.05 and 0.25 solar metallicity). In \citet{2016PhRvD..93h4029R}, $N_{\rm BBH,esc} (t_{\rm GW}<10~{\rm Gyr})$ is the number of BBH merged by the present day.}
\label{tab:Results}
\end{table*}

In Table \ref{tab:Results}, we summarized the number of BBHs merged within 10\,Gyr (hereafter, merging BBHs), and they are 36 and 15 for models A and B, respectively. The number of merging BBHs per solar mass is calculated to be $4.0\times10^{-5}$ and $1.7\times10^{-5}$ for models A and B, respectively. In Table \ref{tab:Results}, we also summarized the results of \citet{2016PhRvD..93h4029R}, in which a series of Monte-Carlo simulations was performed for cluster models with a mass of $\sim 10^5M_{\odot}$. If we compare their model with an initial half-mass density comparable to ours, the number of merging BBHs is 7.2--$9.2\times10^{-5}M_{\odot}^{-1}$. The number of merging BBHs per solar mass for our models corresponds to $\sim 20$--50\,\% of the case of globular clusters.

\citet{2018MNRAS.481.5123B} also performed $N$-body simulations for star clusters with a mass of $\sim 10^4M_{\odot}$. The number of merging BBHs per cluster in their simulation is 0--4. 
The number of mergers per stellar mass is calculated as $\sim 10^{-4}$, which is an order of magnitude higher than ours, although the number of runs are quit few in their paper and the cluster density of their model is an order of magnitude lower than ours. Another possible reason is that they include post-Newtonian terms during their $N$-body simulations. 

Thus, the merging BBHs originated from open clusters is not negligible, if the mass fraction of low-mass star clusters is enough large compared with that of globular clusters. In the following, we estimate the merger rate density of BBHs originated from open clusters.

\begin{figure*}
\begin{center}
\includegraphics[width=0.9\textwidth]{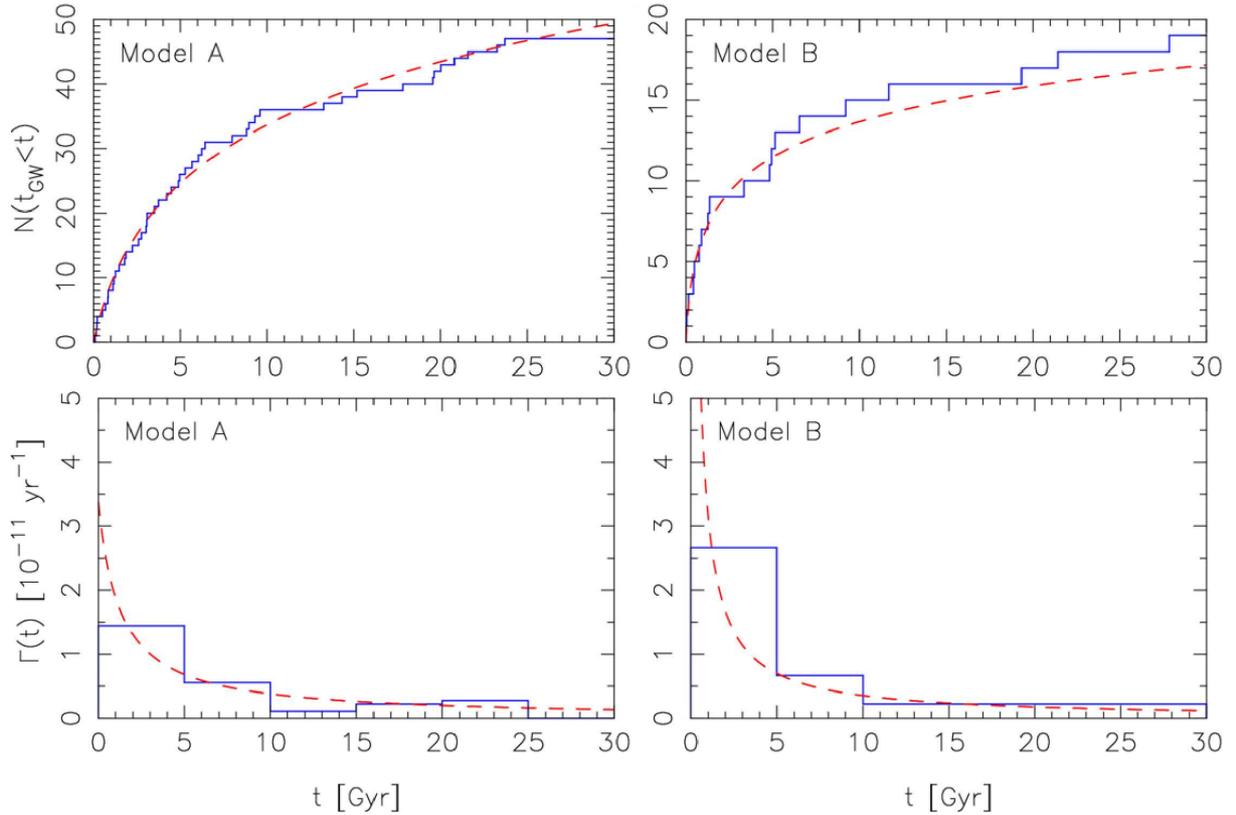}
\end{center}
\caption{Delayed time distribution of merged BBHs. Top panels show the cumulative number of merged BBHs as a function of time for Models A and B. The solid lines show the results of our simulation and the dashed lines show the fitted lines with equation \ref{eq:Nmerge}. Bottom panels show the merger rate per cluster for Models A and B. 
}
\label{fig:DTD}
\end{figure*}

Top panels of Figure \ref{fig:DTD} show the cumulative number of merger ($N_{\rm merger,cum}$) as a function of time for Model A and B. Solid lines show the results of our simulation. Dashed lines are applied fitting to the solid lines using following function, 
\begin{equation}
  N_{\rm merger,cum}(t) = N_0 \ln{\left(\frac{t}{\tau}+1\right)},
  \label{eq:Nmerge}
\end{equation}
where $N_0$ and $\tau$ are fitting parameters. As a results of fitting, these parameters are determined as
\begin{equation}
  \left.
  \begin{aligned}
    &N_0 = 15.4 \\
    &\frac{1}{\tau} = 0.79~{\rm Gyr^{-1}}\\
  \end{aligned}
  \right\}
  \qquad \text{for Model A,}
\end{equation}
\begin{equation}
  \left.
  \begin{aligned}
    &N_0 = 3.2 \\
    &\frac{1}{\tau} = 6.78~{\rm Gyr^{-1}}\\
  \end{aligned}
  \right\}
  \qquad \text{for Model B.}
\end{equation}
The merger rates per cluster are
\begin{equation}
  \Gamma_{\rm }(t) = \frac{\mathrm{d}N_{\rm merger,cum}(t)}{\mathrm{d}t} \times \frac{1}{N_{\rm run}},
\end{equation}
therefore,
\begin{equation}
  \Gamma_{\rm A}(t) \sim 3.4\times10^{-11}\left[0.79\left(\frac{t}{\rm Gyr}\right)+1\right]^{-1}~{\rm yr^{-1}},
\end{equation}
and
\begin{equation}
  \Gamma_{\rm B}(t) \sim 2.4\times10^{-10}\left[6.78\left(\frac{t}{\rm Gyr}\right)+1\right]^{-1}~{\rm yr^{-1}},
\end{equation}
for Models A and B, respectively.

Using the obtained delayed time distributions, we calculate the merger rate density in the local universe. 
We assume that the number density of star clusters ($n_{\rm cl}$) follows a mass function with a power-low of $-2$, i.e., 
star clusters with a mass of $10^3$--$10^4M_{\odot}$ are 100 times more populous than those with a mass of $10^5$--$10^6M_{\odot}$. Since the globular cluster survive until the present day, the number density of globular clusters ($n_{\rm GC}$) is observed as $8.4h^3$\,Mpc$^{-3}$ \citep{2000ApJ...528L..17P} and $0.77$\,Mpc$^{-3}$ \citep{2015PhRvL.115e1101R}. For comparison with previous studies we adopt $n_{\rm GC}=0.77$\,Mpc$^{-3}$. The typical mass of golubular clusters is $10^5$--$10^6M_{\odot}$, we can assume that $n_{\rm cl}=100n_{\rm GC}$ for our models. If we assume that all clusters are $10~{\rm Gyr}$ old, the merger rate density in the local universe is described as $\Gamma(t=10~{\rm Gyr})\times n_{\rm cl}$. The resulting merger rate density are $0.29$ and $0.27~{\rm yr^{-1}Gpc^{-3}}$, for models A and B, respectively.

On the other hand, the merger rate density of globular clusters was estimated to be $\sim 5$\,yr$^{-1}$\,Gpc$^{-3}$ in the local universe (redshift $z<0.5$) \citep{2016ApJ...824L...8R,2016PhRvD..93h4029R}. Therefore, the merger rate density of with a mass of $10^3$--$10^4 M_{\odot}$ corresponds to $\sim 5$\,\% of globular clusters, although our method for estimating the merger rate density may be different from that in \citet{2016ApJ...824L...8R,2016PhRvD..93h4029R}. 

\subsection{Mass ratio of merging black hole binaries}
\begin{figure*}
\begin{center}
\includegraphics[width=0.85\textwidth]{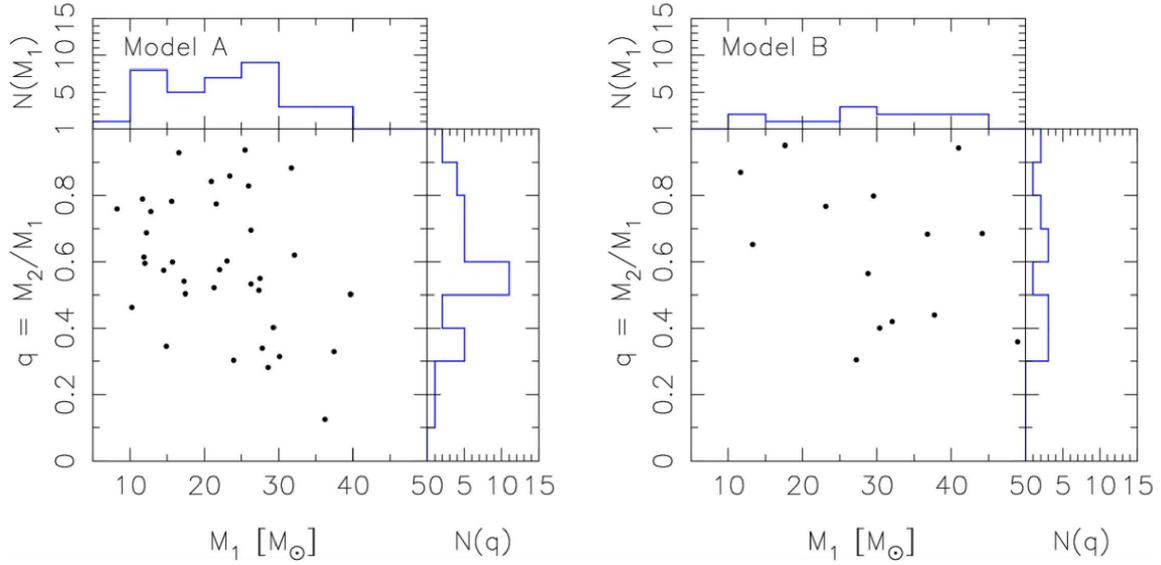}
\end{center}
\caption{The distribution of the mass ration and the primary mass of BBHs whose merger time is less than 10\,Gyr for Models A and B.
}
\label{fig:M1_q}
\end{figure*}

In Figure \ref{fig:M1_q}, we present the mass ratio ($q$) of merging BBHs. In previous studies of merging BBHs ejected from globular clusters, the mass ratio is distributed in $q>0.5$ \citep{2013MNRAS.435.1358T,2016MNRAS.463.2109R}.
In Model A, however, the mass ratio $q$ decreases with the primary mass ($M_{1}$) increases. This is due to the low cluster mass of Model A. We allowed to assigned a massive mass randomly up to $150M_{\odot}$ following an IMF. Because of this setting, the maximum mass ratio of BBHs, which can be formed in the cluster, is sometimes less than 0.5, even if the most and second massive stars form a binary.

\section{Conclusion}
We performed a series of direct $N$-body simulations of open clusters with a mass of $10^3$--$10^4 M_{\odot}$. The number of mergers per solar mass obtained from our simulations are $4.0\times10^{-5}$ and $1.7\times10^{-5}M_{\odot}^{-1}$ for star clusters with a mass of $2.5\times 10^{3}$ and $1.0\times10^4M_{\odot}$, respectively. This corresponds to $\sim 20$--$50$\,\% of the number of merging BBHs per solar mass originated from globular clusters.

The mergers per solar mass increased as the cluster mass decreases. This is because in less massive clusters, core-collapse time is shorter, and therefore massive main-sequence binaries can form before they evolve to BHs. These binaries experience stable and unstable mass transfer evolution and form hard binaries with nearly zero eccentricities, which can merge within the Hubble time. The fraction of merging BBHs through this path becomes larger as the cluster mass decreases. 

Assuming that the initial mass function of star clusters follows a power-law of $-2$, the contribution of merging BBHs originated from open clusters is estimated to be $\sim 0.3$\,yr$^{-1}$\,Gpc$^{-3}$.

\section*{Acknowledgements}

This work was supported by JSPS KAKENHI Grant Number 17H6360 and 16K17656.
Numerical calculations in this paper is supported by Initiative on Promotion of Supercomputing for Young or Women Researchers, Supercomputing Division, Information Technology Center, The University of Tokyo.






\bsp	
\label{lastpage}
\end{document}